\documentclass{article}
\usepackage{amsmath}
\usepackage{amssymb}
\usepackage{amsthm}
\usepackage{graphicx}
\usepackage{psfrag}
\usepackage[hang, nooneline]{subfigure}

\author{Natascha Riahi \footnote{e-mail address: natascha.riahi@gmx.at}\\ University of Vienna, Faculty of Physics, Gravitational Physics
\\ Boltzmanng. 5, 
1090 Vienna, Austria}
\title{The evolution of quantal uncertainties in at
most quadratic potentials}
\begin{document}
\maketitle
\begin{abstract}
We investigate the time evolution of momentum and position uncertainties
for wave packets of arbitrary shape in at most quadratic potentials.
We consider all possible cases of potentials and initial conditions.
Doing so we see that the mixed uncertainty and for Gaussian
wave packets moreover the chirp are convenient and important tools to identify
the characteristic features of uncertainty dynamics. Special attention is given to the spreading and narrowing of wave packets.
\end{abstract}
\section{Introduction}

The fact that the time evolution of position and momentum expectation 
values follows Hamilton's equations of motion, if the potential is at most
quadratic, turns up in any quantum mechanics course. In contrast the 
characteristics of the corresponding dynamics of quantal uncertainties are in general not presented in courses and textbooks - apart from the folk theorem about the spreading of the wave packet for the free particle which only holds under certain conditions. 

Therefore a self-contained derivation and discussion of the behavior of quantal uncertainties for this special class of  potentials seems to be useful for the student to complete his or her basic knowledge of quantum mechanics as well as for
the researcher who wants to start work in some area of quantum dynamics.

This article contains a complete analysis of the dynamics of position and momentum
uncertainties of wave packets moving in at most quadratic potentials according to the Schr\"odinger equation. Some of the results we produce are already given in \cite{He,Sty,An}. But we obtain them without restricting to certain wave packets and we consider the whole range of initial values, identifying all possible kinds of dynamic behavior. We especially investigate which initial values lead to spreading and which to narrowing of wave packets. 

We find that for potentials which allow unbounded motion all wave packets will spread after a certain time. It turns out that the initial values of position and momentum uncertainties are not enough  to prescribe the time evolution of these two quantities, but together with the initial value of the so-called mixed uncertainty the dynamics of all three quantities is determined. We explain the relation of the mixed uncertainty and the chirp, which is defined for Gaussian wave packets, and also investigate the time evolution of both quantities.

This article is structured as follows: Section~\ref{Equ} contains the derivation of the differential equations which govern the uncertainty dynamics. We further introduce the mixed uncertainty and the chirp and identify the constants of motion. In section~\ref{Free} the reader will find the uncertainty analysis for at most linear potentials. Section~\ref{HO} covers the case of the harmonic oscillator, and finally the unstable harmonic oscillator is treated in section~\ref{uHO}.

\section{Time evolution of uncertainties}
\label{Equ}
We investigate the dynamics of wave packets subject to a Hamiltonian 
\begin{equation}
H=\frac{\hat p^2}{2 m}+V(x)
\end{equation}
where the potential V(x) has the form
\begin{equation}
\label{quadratisch}
V(x)=A x^2 +B x +C \qquad \mbox{with} \qquad A\,,\,B 
,\,C\,\epsilon \,\mathbb{R} \quad .
\end{equation}
The evolution of the expectation value of any time-independent
operator $ \hat O$ is given by
\begin{equation}
\frac{d}{dt} \left\langle \hat O \right\rangle=-\frac{i}{\hbar}
\left\langle \left [ \hat O, \hat H \right ] \right\rangle \quad.
\end{equation}
Therefore the position and momentum expectation values evolve according to Newton's equations
\begin{align}
\frac{d}{dt} \left\langle \hat x \right\rangle=
\frac{\left\langle \hat p \right\rangle}{m}\, , \qquad
\frac{d}{dt} \left\langle \hat p \right\rangle=
-\left\langle \frac{\partial V(x)}{\partial x} \right\rangle=
-2 A \left\langle \hat x \right\rangle-B \quad.
\end{align}
For the first and second time derivative of the position and momentum uncertainties 
\begin{equation}
(\Delta x)^2=\left\langle \hat x^2\right\rangle-\left\langle \hat x\right\rangle^2 \,,
\qquad
(\Delta p)^2=\left\langle \hat p^2\right\rangle-\left\langle \hat p\right\rangle^2
\end {equation}
we find in an arbitrary potential, using the abbreviations $F=-\frac{\partial V(x)}{\partial x}$
and $F'=-\frac{\partial^2 V(x)}{\partial x^2}$,
\begin{subequations}
\label{deltapa}
\begin{align}
\label{deltaxpa}
&\frac{d}{dt}  (\Delta x)^2=
\frac{1}{m} \left\langle 
\hat x \hat p + \hat p\, \hat x - 
2\left\langle \hat x \right\rangle \left\langle \hat p\right\rangle
\right\rangle\, , \\
\label{deltappa}
&\frac{d}{dt}  (\Delta p)^2=
\left\langle 
\left(\hat p-\left\langle \hat p\right\rangle\right)F\right\rangle+
\left\langle F \left(\hat p-\left\langle \hat p\,\right\rangle\right)
\right\rangle\, ,\\
\label{deltaxppa}
&\frac{d^2}{dt^2}  (\Delta x)^2 =
\frac{2}{m^2}(\Delta p)^2+
\frac{2}{m}\left\langle 
\left(\hat x-\left\langle \hat x\right\rangle\right)F
\right\rangle\, , \\
\label{deltapppa}
&\frac{d^2}{dt^2}  (\Delta p)^2 =
\frac{1}{2m} \left(\left\langle 
\hat p ^2 F'+2 \hat p F' \hat p +F' \hat p^2\right\rangle-
2\left \langle \hat p F'+F' \hat p\right\rangle \left\langle \hat p \right \rangle \right)  \nonumber \\
& \qquad +
2 \left(
\left\langle F^2 \right\rangle -\left\langle F \right\rangle^2
\right) \quad.
\end{align}
\end{subequations}
Equation (\ref{deltaxpa}) does not require any information about the potential.
We will call the quantity on the right hand side, which characterizes the first derivative of the position uncertainty, the mixed uncertainty:

\begin{equation}
\label{mixed}
 \Delta_{x p}=\frac{1}{2}\left\langle\hat x \hat p + \hat p\, \hat x - 
2\left\langle \hat x \right\rangle \left\langle \hat p\right\rangle
\right\rangle \quad.
\end{equation}
 
The mixed uncertainty is always real, but not necessarily positive, which distinguishes it from the position and momentum uncertainty.
 In accordance with (\ref{deltaxpa}) the sign of the mixed uncertainty determines if the wave packet
is going to spread or narrow at a given time. This particular statement is true  for
an arbitrary potential since we have not used the special form of the potential (\ref{quadratisch}) for the calculation of (\ref{deltapa}).
Moreover $(\Delta x)^2,\,(\Delta p)^2$ and $\Delta_{xp}$  satisfy the inequality
\begin{equation}
\label{GenU}
(\Delta x)^2 (\Delta p)^2 -(\Delta_{xp})^2 \geq \frac{1}{4} \hbar^2 \, ,
\end{equation}
which is stronger than Heisenberg's uncertainty inequality \cite{An}. For
Gaussian wave packets the inequality (\ref{GenU}) becomes an equality (for more about Gauss functions see below).

If we insert the expression of the mixed uncertainty (\ref{mixed}) and use
the special form of the potential (\ref{quadratisch}) the equations 
(\ref{deltapa}) become
\begin{subequations}
\label{deltap}
\begin{align}
\label{deltaxp}
&\frac{d}{dt}  (\Delta x)^2=
\frac{2}{m} \Delta_{xp}\quad ,\\
\label{deltapp}
&\frac{d}{dt}  (\Delta p)^2=
-4 A \Delta_{xp}\quad , \\
\label{deltaxpp}
&\frac{d^2}{dt^2}  (\Delta x)^2 =
\frac{2}{m^2}(\Delta p)^2-\frac{4 A}{m}(\Delta x)^2\, , \\
\label{deltappp}
&\frac{d^2}{dt^2}  (\Delta p)^2 =
8 A^2(\Delta x)^2-\frac{4 A}{m}(\Delta p)^2 \,.
\end{align}
\end{subequations}

The equations (\ref{deltaxpp}) and (\ref{deltappp}) represent a system of
two linear second order differential equations which has a unique solution for 
fixed initial values of $(\Delta x)^2,\,(\Delta p)^2,\,\frac{d}{dt} (\Delta x)^2$ and $\frac{d}{dt}  (\Delta p)^2$. In order to fulfill equation (\ref{deltaxp})  the initial value of $\frac{d}{dt}(\Delta x)^2$ has to be chosen in accordance with the initial mixed uncertainty. But if finally equation (\ref{deltapp}) should also be satisfied, the relation
\begin{equation}
\label{constraint}
\frac{d}{dt} \left(2 m A (\Delta x)^2+(\Delta p)^2\right)=0
\end{equation}
must hold for all times. Since a closer look at the system (\ref{deltaxpp},\ref{deltappp}) reveals
\begin{equation}
\frac{d^2}{dt^2} \left(2 m A (\Delta x)^2+(\Delta p)^2\right)=0
\end{equation}
we see that the constraint equation(\ref{constraint}) is fulfilled during the time evolution of the system (\ref{deltaxpp},\ref{deltappp}) if it is only satisfied
at $t=0$. So we get a unique solution for (\ref{deltap}) if we solve (\ref{deltaxpp},\ref{deltappp}) and choose the initial values according to  (\ref{deltaxp}, \ref{deltapp}). 

The system (\ref{deltap}) has two constants of motions, namely
\begin{equation}
\label{Konstante}
K=(\Delta p)^2+2 m A(\Delta x)^2 \quad \mbox{and} \quad
U=(\Delta x)^2 (\Delta p)^2 -(\Delta_{xp})^2 \quad.
\end{equation}

The conservation of $K$ is equivalent to the constraint equation (\ref{constraint}). $U$ is just the left side of the inequality (\ref{GenU}). The conservation of $U$  can be derived carrying out the time derivative of $U$ and using (\ref{deltaxp},\ref{deltapp},\ref{deltaxpp}). If the mixed uncertainty takes the 
value zero, which is in particular the case when $(\Delta x)^2$ or $(\Delta p)^2$ have an extreme value (see (\ref{deltaxp},\,\ref{deltapp})), the uncertainty product P takes the value of its global minimum 
\begin{equation}
\label{result}
P=(\Delta x)^2(\Delta p)^2=P_{min}=U \,.
\end{equation}

The initial values for the three uncertainties can in general be chosen arbitrarily,
as long as the generalized uncertainty relation (\ref{GenU}) is satisfied. But of course all three quantities are determined, if a certain wave packet is given.

In particular, a Gaussian wave packet of the shape

\begin{align}
\label{Gaussian}
&\Psi(x)=\frac{1}{(\pi \alpha)^{1/4}}e^{-\frac{(x-x_{0})^2}{2 \alpha }
+\frac{i p_{0}(x-x_{0})}{\hbar} +i a (x-x_{0})^2 +i \varphi} \\
& \mbox{with}\, \alpha\,>\,0 \quad\mbox{and}\quad
 x_{0},p_{0},a, \varphi \,\epsilon \,\mathbb{R} \nonumber
\end{align}
has the properties
\begin{align}
\label{properties}
 &\left\langle \hat x \right\rangle=x_{0} \,,
\quad &(\Delta x)^2 =\frac{\alpha}{2}, \quad \Delta_{xp}=2 a \hbar\,(\Delta x)^2\,,\\
&\left\langle \hat p \right\rangle=p_{0}\,,\quad 
&(\Delta p)^2 =\frac{\hbar^2}{4 (\Delta x)^2} +4 a^2 \hbar^2 (\Delta x)^2
.\nonumber
\end{align}
So we see that the mixed uncertainty is proportional to the parameter a, which is also called chirp since its effect on the wavefunction looks
 like the modulation of the frequency of a signal characterized by $e^{\frac{i p_{0}(x-x_{0})}{\hbar}}$ (for more details see \cite{He}). If $a=0$ the 
wave packet fulfills Heisenberg's uncertainty relation exactly so that $(\Delta p)^2$ is determined by $(\Delta x)^2$. The freedom to 
choose $a$ in the range $\left(-\infty,\infty\right)$ implies the freedom to choose $(\Delta p)^2$ independently of 
$(\Delta x)^2$ as long as Heisenberg's uncertainty is not violated.
$(\Delta x)^2$ and $(\Delta p)^2$ determine the mixed uncertainty  according to  (\ref{properties}), which finally leads to:
\begin{equation}
\label{GaussianU}
U=(\Delta x)^2 (\Delta p)^2 -(\Delta_{xp})^2= \frac{1}{4} \hbar^2 \quad.
\end{equation}
So we see that the inequality (\ref{GenU}) becomes an equality for Gaussian wave packets. This property will be preserved if the Gaussian wave packet is 
subject to a time evolution governed by an at most quadratic Hamiltonian, since $U$ (\ref{Konstante}) is a constant of motion. This result is in accordance with 
the often used fact that a Gaussian wave packet of the form (\ref{Gaussian}) evolves as a Gaussian if the potential is at most 
quadratic (for a proof see for instance \cite{He}). Therefore the time evolution of a Gaussian wave packet (\ref{Gaussian}) subject to an at most 
quadratic Hamiltonian is described by the functions $\alpha(t),a(t),x_{0}(t),p_{0}(t),\varphi(t)$ which represent the time evolution of the characteristic parameters. 
The results for $\alpha(t),a(t)$ can be determined by (\ref{properties}), if the solutions of (\ref{deltap}) are given.

All Gaussian wave packets with the property
\begin{equation}
\Delta_{xp}=a=0
\end{equation}
fulfill Heisenberg's uncertainty relation exactly and are therefore called minimum uncertainty states.

\section{Free particle and linear potential}
\label{Free}

Since the parameters $B$ and $C$ do not appear in the equations (\ref{deltap}), the dynamics of uncertainties is the same for all potentials of the form
\[
V(x)=B\,x +C \quad.
\] 
In this case the momentum uncertainty is a constant of motion,
\[
K= (\Delta p)^2 = (\Delta p_{0})^2\quad.
\] 
With the initial values $(\Delta x)^2_{t=0} =(\Delta x_{0})^2$ and $(\Delta_{xp}) _{t=0} =\Delta^{0}_{xp}$ (\ref{deltaxpp}) and (\ref{deltaxp}) yield
\begin{subequations}
\label{deltaf}
\begin{align}
\label{dxf}
&(\Delta x)^2=(\Delta x_{0})^2+\frac{t^2}{m^2} (\Delta p_{0})^2 +
\frac{2 t}{m} \Delta^{0}_{xp} \\
\label{dxpf}
&\Delta_{xp}=\Delta^{0}_{xp}+\frac{t}{m} (\Delta p_{0})^2
\end{align}
\end{subequations}
The mixed uncertainty is a monotonically increasing function in time.
If $\Delta^{0}_{xp} \geq 0$, $(\Delta x)^2$ is monotonically increasing. If $\Delta^{0}_{xp} < 0$, $(\Delta x)^2$ is monotonically decreasing until it reaches
its minimum at
\begin{equation}
\label{Tmin}
T=- \frac{m \Delta^{0}_{xp}}{(\Delta p_{0})^2} \quad,
\end{equation}

when the mixed uncertainty is zero. At $t=T$ the uncertainty product also has its minimum value and the minimum value of the position uncertainty reads in accordance with (\ref{result})

\begin{equation} 
\label{result1}
(\Delta x)^2 (T)=(\Delta x)^2 _{min}=U/(\Delta p_{0})^2 \quad.
\end{equation}

A wave packet that starts  with a negative mixed uncertainty is going to narrow until $t=T$. If we insert $ \Delta_{xp}^{0} = -\sqrt{(\Delta x_{0})^2(\Delta p_{0})^2-U}$ in (\ref{Tmin}), we find
\[
T=\frac{m \Delta x_{0}}{(\Delta p_{0})} \sqrt{1-\frac{U}{(\Delta x_{0})^2(\Delta p_{0})^2}} \quad.
\]
This means that for any $U\geq \hbar^2/4$ and for any initial uncertainty product
$(\Delta x_{0})^2(\Delta p_{0})^2> U$, $T$ can take any given value in the interval $(0,\infty)$. So it may take arbitrary long time until the wave packet stops narrowing and starts to spread which is the behavior commonly expected from a free particle wave packet. But of course all free particle wave packets are going to spread for $t\rightarrow\infty$. 

\begin{figure}[htp]
\psfrag{t}[cc][cc]{$\tau$} 
\psfrag{aDx2}[cl][cl]{$a , (\Delta x)^2$} 
\psfrag{tmin}[cc][bc]{$ \tau_{min}$} 
\psfrag{tmax}[cc][bc]{$ \tau_{max}$} 
\psfrag{Dx2min}[cc][bl][0.8]{$(\Delta x)^2_{min}$} 
\psfrag{zmalmin}[bc][tl][0.8]{$2 (\Delta x)^2_{min}$} 
 \includegraphics[width=0.45\textwidth]{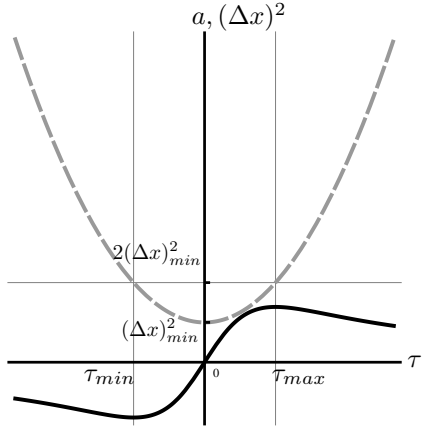} 
\caption{The time dependence of position uncertainty and chirp for a Gaussian wave packet in an at most linear potential.
The chirp $a(\tau)$ has a maximum (minimum) at $\tau_{max}$ ($\tau_{min}$) when $(\Delta x)^2$ (dashed line)
reaches twice its minimum value $(\Delta x)^2_{min}$ . }
  \label{FreeP}
\end{figure}

If we make the replacement $\tau=t-T$ in (\ref{deltaf}), we can express all possible histories of the uncertainties in terms of the constants of motion:
\begin{subequations}
\label{Orbitsf}
\begin{align}
&(\Delta x)^2=(\Delta p_{0})^2 \frac{\tau^2}{m^2}+U/(\Delta p_{0})^2\,, \\
&\Delta _{xp}=(\Delta p_{0})^2 \frac{\tau}{m} \quad.
\end{align}
\end{subequations}

This reveals that all possible histories of the position uncertainty consist of a narrowing and a spreading part. Furthermore the mixed uncertainty always goes from $-\infty$ to $\infty$.
If the initial wavefunction is a Gaussian wave packet ($U=\frac{\hbar^2}{4}$), the equations (\ref{properties}, \ref{GaussianU}) and
(\ref{Orbitsf}) yield for the chirp 
\[
a(\tau)=\frac{ \Delta_{xp}}{2 \hbar (\Delta x)^2 }=
\left(\frac{m}{2\hbar} \right) \frac{\tau}{\tau^2+\frac{m^2 \hbar^2}{4( \Delta p_{0})^4}} \quad.
\]
So $a$ has a maximum at the time $\tau_{max}= \hbar\, m/(2 (\Delta p_{0})^2) $  and a minimum at the time $\tau_{min}=-\tau_{max}$ and takes the values $a_{max}=(\Delta p_{0})^2/(2 \hbar ^2 )$
and $a_{min}=-(\Delta p_{0})^2/(2 \hbar ^2 )$ (see also figure \ref{FreeP}). If $K=(\Delta p_{0})^2 $ tends to infinity the time interval between the maximal and the minimal chirp tends to zero. Furthermore we see that the chirp always approaches zero if $\tau$ tends to infinity.

\section{Harmonic oscillator}
\label{HO}

We will now consider the dynamics of uncertainties in the potential of a harmonic oscillator with the particle mass $m$ and the angular frequency $\omega$. 
If we insert $A=\frac{m \omega ^2}{2}$ in (\ref{deltap}) and use the variables

\begin{align}
\label{variablesO}
K=(\Delta p)^2+ m^2 \omega^2 (\Delta x)^2  \qquad
Z=(\Delta p)^2- m^2 \omega^2 (\Delta x)^2
\end{align}

instead of $(\Delta x)^2$ and $(\Delta p)^2$, we get the system of equations
\begin{align}
\label{Op}
&\frac{d}{dt}  K=0
&\frac{d}{dt} Z=
-4 m \omega^2 \Delta_{xp} \\
&\frac{d^2}{dt^2} K =0 
&\frac{d^2}{dt^2}  Z = - 4 \omega^2 Z \nonumber  \quad.
\end{align}

The solution of (\ref{Op}) reads
\begin{align}
\label{ZK}
K=const. 
\qquad & Z=Z_{0} \, \mbox{cos}(2 \omega t)-
\Delta^{0}_{xp} \,2 m \omega\, \mbox{sin}(2 \omega t) \\
\qquad & \Delta_{xp}=\Delta^{0}_{xp}\, \mbox{cos}(2 \omega t)
+\frac{Z_{0}}{2 m \omega }\,\mbox{sin}(2 \omega t) \nonumber
\end{align} 

with the initial values $Z_{0}=Z_{t=0}$ and $\Delta^{0}_{xp}=(\Delta_{xp}) _{t=0} $. So we see that the variable $Z$ and the mixed uncertainty $\Delta_{xp}$ oscillate twice as fast as the expectation values of the classical phase space variables  and so will the uncertainties $(\Delta x)^2$ and $(\Delta p)^2$.
For the special initial conditions $Z_{0}=\Delta_{xp}^{0}=0$, the mixed uncertainty and the variable $Z$ stay zero for all times. In all other cases, $Z$ has one maximum
during the oscillation period $[0,\frac{\pi}{\omega}]$.
We find for the maximal value of $Z$

\begin{equation}
Z_{max}=(Z_{0}^2+4 m^2 \omega^2(\Delta^{0}_{xp})^2)^{1/2} 
  =(K^2-4 m^2 \omega^2 U)^{1/2} 
\end{equation}

where we have used the definitions of $K, Z$ (\ref{variablesO}) and $U$ (\ref{Konstante}) in the last step. At the time $T$ when $Z$ takes its maximum value, the mixed uncertainty reaches zero.
We will now choose $T$ as starting point and therefore make the replacement $\tau=t-T$.
So we find for the time evolution of $Z$  
\[
Z=(K^2-4 m^2 \omega^2 U)^{1/2} \mbox{cos}(2 \omega \tau) \,,
\]
which yields for the uncertainties
\begin{align}
\label{Oszi}
&\Delta_{xp}=\frac{(K^2-4 m^2 \omega^2 U)^{1/2}}{2 m \omega}\,
\mbox{sin} (2 \omega \tau)\,, \\
&(\Delta x)^2=\frac{1}{2 m^2 \omega^2} 
\left(K- 
(K^2-4 m^2 \omega^2 U)^{1/2}\,
\mbox{cos} (2 \omega \tau)\right) \,,\nonumber \\
&(\Delta p)^2=\frac{1}{2} 
\left(K+
(K^2-4 m^2 \omega^2 U)^{1/2}\,
\mbox{cos} (2 \omega \tau)\right) \nonumber \quad.
\end{align}

The definitions of $U$ and $K$ imply 

\begin{equation}
\label{inequality}
K^2-4 m^2 \omega^2 U\,\geq\,0 \qquad,
\end{equation}

since

\begin{align}
\label{Pinequality}
K^2-4 m^2 \omega^2 U\,&=\,((\Delta p)^2+m^2\omega^2(\Delta x)^2)^2-
4 m^2\omega^2((\Delta p)^2(\Delta x)^2-(\Delta_{xp})^2) \nonumber
\\& \geq\,
((\Delta p)^2+m^2\omega^2(\Delta x)^2)^2-
4 m^2\omega^2(\Delta p)^2(\Delta x)^2 \\
\quad &=((\Delta p)^2-m^2\omega^2(\Delta x)^2)^2
\,\geq\,0 \quad.\nonumber
\end{align}

We see that the initial condition 
\begin{equation}
\label{inicon}
K^2-4 m^2 \omega^2 U=0 \, 
\end{equation}
 is equivalent to 
 \begin{equation}
 \Delta_{xp}^{0}=0 \quad \mbox{and} \quad
 (\Delta p_{0})^2=m^2\omega^2 (\Delta x_{0})^2 \quad.
 \end{equation}
For this special initial condition the mixed uncertainty continues to be zero and the other two uncertainties remain constant. Furthermore the uncertainty product fulfills
\begin{equation}
\label{result2a}
(\Delta x)^2 (\Delta p)^2=U \quad \forall\, \tau\, \quad .
\end{equation}

\begin{figure}[htp]
\centering
\subfigure[The variable $\tilde a \equiv \frac{a  2 \hbar}{m\omega}$ for $k=1.2$ (light-gray), 
$k=1.5$ (gray) and $k=2.5$. The time between maximum and minimum becomes shorter if k becomes bigger.]
{\label{Oszia} 
\psfrag{a}{$\tilde a$}
\psfrag{tau}{$\tau$}
\psfrag{Pd4o}[cc][cl][0.6]{$\pi/(4 \omega)$}
\psfrag{Pd2o}[cc][cl][0.6]{$\pi/(2 \omega)$}
\psfrag{DPd4o}[cc][cc][0.6]{$3 \pi/(4 \omega)$}
\includegraphics[width=0.45\textwidth]{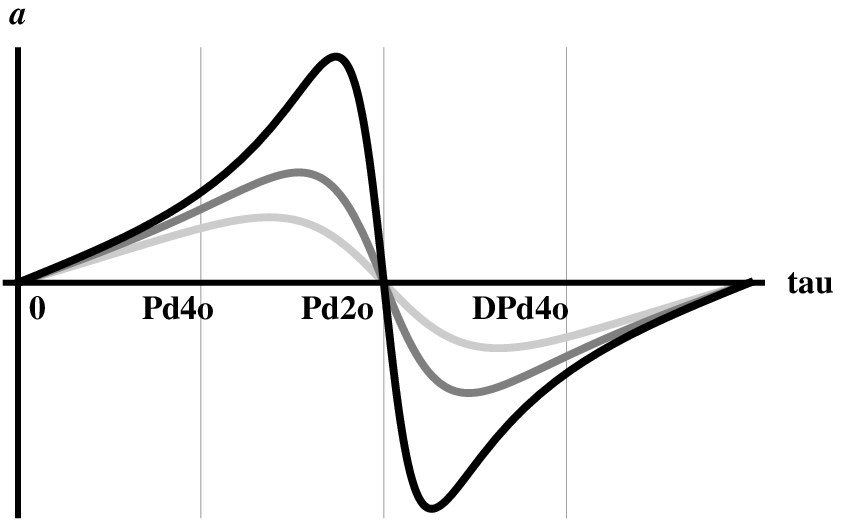}}
\subfigure[The oscillations of $(\Delta X)^2\equiv(\Delta x)^2 \frac{2 m\omega}{\hbar}$,
  $(\Delta P)^2\equiv(\Delta p)^2 \frac{2} {m \omega \hbar}$ (light-gray) and
  $(\Delta_{XP})\equiv(\Delta_{xp}) \frac{2} {\hbar}$ (gray)  for $k=1.2$ ]{\label{Oszib} 
\psfrag{tau}{$\tau$}
\psfrag{DX2DP2DXP}[cc][cc]{$(\Delta X)^2, (\Delta P)^2, \Delta_{XP}$}
\psfrag{tau}{$\tau $}
\psfrag{Pd4o}[cc][cc][0.6]{$\pi/ (4 \omega)$}
\psfrag{Pd2o}[cc][cc][0.6]{$\pi/ (2 \omega)$}
\psfrag{DPd4o}[cc][cc][0.6]{$3 \pi /4 \omega$}
\psfrag{Pdo}[cc][cc][0.6]{$\pi /\omega$}
\includegraphics[width=0.45\textwidth]{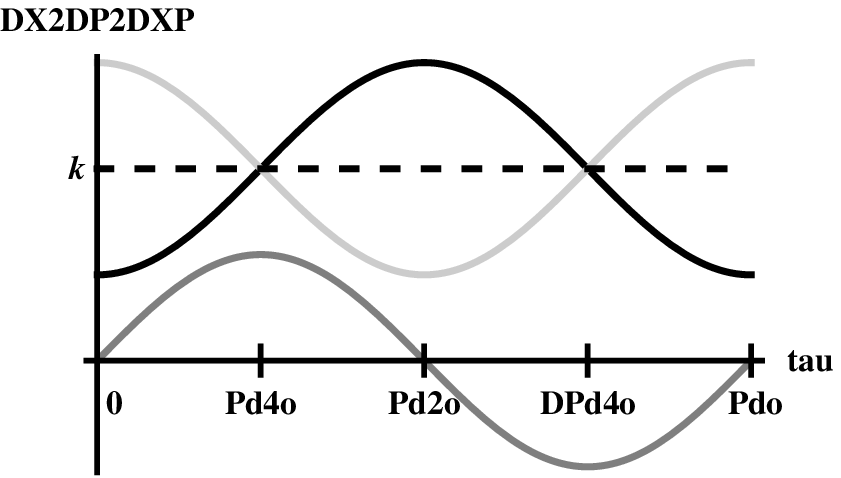}}
\caption{The evolution of chirp and uncertainties for a Gaussian wave packet in a harmonic oscillator potential during one full period.}
\label{Osziab}
\end{figure}

For all other possible initial values, $(\Delta p)^2$ and $(\Delta x)^2$ oscillate around the values $\frac{K}{2}$ and $\frac{K}{2 m^2 \omega^2}$. The mixed uncertainty  oscillates around zero and takes $\pm \frac{(K^2-4 m^2 \omega^2 U)^{1/2}}{2 m \omega}$ as maximum and minimum value, respectively.
We find that the position and the momentum uncertainty always have an extremum at the instants when the mixed uncertainty is zero. In particular $(\Delta x)^2$ takes a maximum and $(\Delta p)^2$ a minimum for $\tau=\pi/(2 \omega)$ during one period and vice versa for $\tau=0,\pi/\omega$ (see also figure (\ref{Oszib}). In accord with (\ref{result}) we get

\begin{align}
\label{result2b}
&(\Delta x)^2_{min} (\Delta p)^2_{max}=U \, &\mbox{for}\quad &\tau=0,\pi/\omega \\
&(\Delta x)^2_{max} (\Delta p)^2_{min}=U \, &\mbox{for}\quad &\tau=\pi/(2\omega)
 \nonumber \quad.
\end{align}
Note that we can choose $\Delta_{xp}^{0}$  and one of the other two initial values (for instance $(\Delta x_{0})^2$) arbitrarily. But if we want to get a solution with constant uncertainties, we can not take an arbitrary value for $(\Delta x_{0})^2$. For in this case the initial momentum uncertainty would have to be $(\Delta p_{0})^2=m^2\omega^2 (\Delta x_{0})^2$, which means that the uncertainty relation requires 

\begin{subequations}
\label{DeltaMin}
\begin{equation}
\label{DeltaMin1}
(\Delta x_{0})^2\,\geq\, \frac{\hbar}{2 m \omega} \quad.
\end{equation}

Likewise the initial momentum uncertainties that provide non-oscillatory solutions have to fulfill

\begin{equation}
\label{DeltaMin2}
(\Delta p_{0})^2\,\geq\, \frac{\hbar m \omega}{2} \quad.
\end{equation}
\end{subequations}

If the initial wave function is a Gaussian wave packet $(U=\frac{\hbar^2}{4})$ the inequality (\ref{inequality}) reads $K\,\geq\,m \omega \hbar$. Using  the abbreviation $k=\frac{K}{m \omega \hbar}$ we find for the uncertainties (\ref{Oszi}) and the chirp (\ref{properties})

\begin{align}
\label{OsziGauss}
&\Delta_{xp}=\frac{\hbar}{2}(k^2-1)^{1/2}\,
\mbox{sin} (2 \omega \tau)\,, \\
&(\Delta x)^2=\frac{\hbar}{2 m \omega}
(k-(k^2-1)^{1/2}\,
\mbox{cos} (2 \omega \tau))\nonumber\,, \\
&(\Delta p)^2=\frac{m \omega \hbar}{2}
(k+(k^2-1)^{1/2}\,
\mbox{cos} (2 \omega \tau))\nonumber\,, \\
&a=\frac{m \omega}{2 \hbar} \,
\frac{\mbox{sin}(2 \omega t)}{\frac{k}{\sqrt{k^2-1}}+\mbox{cos}(2 \omega t)} 
\quad. \nonumber
\end{align}

The chirp oscillates with the frequency $2\omega$ as the other variables. It has a maximum at the time 
\[
\tau_{max}=\frac{\mbox{arcsec} \left(-\frac{k}{\sqrt{k^2-1}}\right)}
{2\omega}
\]
and a minimum at $\tau_{min}=\frac{\pi}{\omega}-\tau_{max}$. For $k\, \epsilon\,
(1,\infty)$, $\tau_{max}$  varies between $\frac{\pi}{4 \omega}$ and $\frac{\pi}{2 \omega}$. If $k$ tends to infinity $\tau_{max}$ and  $\tau_{min}$ approach the center of the interval $(0,\,\frac{\pi}{\omega})$ from opposite directions (see also figure \ref{Oszia}). The extreme values of the chirp are 

\begin{equation}
a_{max}=\frac{m \omega}{2 \hbar} \sqrt{k^2-1} \quad \mbox{and} \quad
a_{min}=-\frac{m \omega}{2 \hbar} \sqrt{k^2-1}\quad.
\end{equation}

Gaussian wave packets with $k=1$ fulfill the condition (\ref{inicon}). Therefore their uncertainties remain constant for all times. Their position and momentum uncertainty take the minimal possible values among 
all solutions with constant uncertainties (\ref{DeltaMin}), namely
\begin{equation}
\label{Coherent}
(\Delta x_{0})^2\,=\, \frac{\hbar}{2 m \omega}\,,\quad
(\Delta p_{0})^2\,=\, \frac{\hbar m \omega}{2}\quad.
\end{equation}
These uncertainties equal the uncertainties of the ground state of the harmonic oscillator.
The states are identical with the coherent states of the 
harmonic oscillator \cite{Ali,Henry}, as can be verified by calculating
\begin{equation}
 \hat A \left.\Psi \right\rangle=
 \frac{1}{\sqrt{2}} \left.(m \omega \hat x + i \hat p)\,|\Psi \right\rangle=
\frac{1}{\sqrt{2}} \left. (m \omega  x_{0} + i p_{0})\, |\Psi \right\rangle\quad,
\end{equation}
using (\ref{Gaussian},\ref{properties}), where $\hat A$ denotes the annihilation operator. These states remain minimal uncertainty states for all times.

Gaussian wave packets, which are minimum uncertainty states but do not fulfill the condition   $(\Delta p_{0})^2=m^2\omega^2 (\Delta x_{0})^2$, are called squeezed states \cite{Henry}, because either the position or the momentum uncertainty are squeezed beyond the initial value for non-oscillating time evolution (\ref{Coherent}). According to (\ref{OsziGauss}) we see, that these states can not remain minimal uncertainty states, since the mixed
uncertainty will always oscillate for $k > 1$, and hence the uncertainty product will also oscillate via equation (\ref{Konstante}).

\section{Unstable harmonic oscillator}
\label{uHO}
The unstable harmonic oscillator is a potential which is quadratic in $x$ and has the shape of a barrier.  
If we insert $A=-\frac{m \omega ^2}{2}$ in (\ref{deltap}) and use the variables

\begin{align}
\label{variablesuO}
K=(\Delta p)^2- m^2 \omega^2 (\Delta x)^2  \qquad
Z=(\Delta p)^2+ m^2 \omega^2 (\Delta x)^2
\end{align}

instead of $(\Delta x)^2$ and $(\Delta p)^2$, we get the system of equations
\begin{align}
\label{uOp}
&\frac{d}{dt}  K=0
&\frac{d}{dt} Z=
4 m \omega^2 \Delta_{xp} \\
&\frac{d^2}{dt^2} K =0 
&\frac{d^2}{dt^2}  Z =  4 \omega^2 Z \nonumber  \quad.
\end{align}

The solution of (\ref{uOp}) reads
\begin{align}
\label{ZKu}
K=const. 
\qquad & Z=Z_{0} \, \mbox{cosh}(2 \omega t)+
\Delta^{0}_{xp} \,2 m \omega\, \mbox{sinh}(2 \omega t) \\
\qquad & \Delta_{xp}=\Delta^{0}_{xp}\, \mbox{cosh}(2 \omega t)
+\frac{Z_{0}}{2 m \omega }\,\mbox{sinh}(2 \omega t) \nonumber
\end{align} 

with the initial values $Z_{0}=Z_{t=0}$ and $\Delta^{0}_{xp}=(\Delta_{xp}) _{t=0} $.

The mixed uncertainty is a monotonically increasing function, as it is already determined by (\ref{uOp}) since
\[
\frac{d \Delta_{xp}}{dt}= \frac{1}{4 m \omega^2}\,\frac{d^2 Z}{dt^2} 
=\frac{Z}{m}\,>0 \quad.
\]

The mixed uncertainty becomes zero at the time
\[
T=\frac{1}{2 \omega} \mbox{artanh} 
\left(-\frac{2 m \omega \Delta^{0}_{xp}}{Z_{0}} \right) \quad.
\]
Starting with a negative initial value,
$ \Delta x_{xp}^{0} = -\sqrt{(\Delta x_{0})^2(\Delta p_{0})^2-U}$, we can write for the time it takes the mixed uncertainty to become zero
\[
T=\frac{1}{2\omega} \mbox{artanh} \left(
\frac{2 m \omega \Delta x_{0}\Delta p_{0}}{Z_{0}}\,
\sqrt{1-\frac{U}{(\Delta x_{0})^2 (\Delta p_{0})^2}} \right) \quad.
\]
The variable $Z$ is bounded from below, since 
\[
Z=(\Delta x)^2+m^2\omega^2(\Delta p)^2\,
\geq\, 2 m\omega (\Delta x)(\Delta p) \quad.
\]
Therefore for any initial uncertainty product $(\Delta x_{0})^2(\Delta p_{0})^2\,>\,U$, $T$ takes values in the interval $(0,T_{max})$ with
\[
T_{max}=\frac{1}{2\omega} \mbox{artanh} \left(\sqrt{1-\frac{U}{(\Delta x_{0})^2(\Delta p_{0})^2}}\,\right) \quad.
\]
So in contrary to the case of the free particle the time until an initially narrowing wave packet starts to spread can only be arbitrary long if the initial uncertainty product $(\Delta x_{0})^2(\Delta p_{0})^2$ takes an arbitrary high value.

At $t=T$ we find for $Z$ 
\[
Z_{t=T}=(Z_{0}^2-4m^2\omega^2(\Delta^{0}_{xp})^2)^{1/2}=(K^2+4 m^2 \omega^2 U)^{1/2}
\]
where we have used the definitions of $K, Z$ (\ref{variablesuO}) and $U$ (\ref{Konstante}) in the last step. If we make the replacement $\tau\,=\,t-T$ the time evolution of $Z$ reads
\[
Z=(K^2+4 m^2 \omega^2 U)^{1/2} \mbox{cosh}(2 \omega \tau)
\]
which yields for the uncertainties
\begin{align}
\label{uOszi}
&\Delta_{xp}=\frac{(K^2+4 m^2 \omega^2 U)^{1/2}}{2 m \omega}\,
\mbox{sinh} (2 \omega \tau)\,, \\
&(\Delta x)^2=\frac{1}{2 m^2 \omega^2} 
\left(
(K^2+4 m^2 \omega^2 U)^{1/2}\,
\mbox{cosh} (2 \omega \tau)-K\right)\,, \nonumber \\
&(\Delta p)^2=\frac{1}{2} 
\left(
(K^2+4 m^2 \omega^2 U)^{1/2}\,
\mbox{cosh} (2 \omega \tau)+K\right) \nonumber \quad.
\end{align}

So all possible histories of the position and the momentum uncertainty consist of a
narrowing and a spreading part. As in the case of the free particle wave packet the mixed uncertainty always goes from $-\infty$ to $\infty$. At $\tau=0$, when the mixed uncertainty vanishes and  the position and the momentum uncertainty have a minimum, the uncertainty product reads in accordance with (\ref{result})

\begin{equation}
\label{result3}
(\Delta x)^2_{min}(\Delta p)^2_{min}=U \quad.
\end{equation}

In the limit $\tau\rightarrow \infty$ the ratio of uncertainties $(\Delta x/\Delta p)$ always tends to $\frac{1}{m\omega}$. Unlike the case of the harmonic oscillator histories with constant uncertainties do not exist. But for $K=0$ the ratio $(\Delta x/\Delta p)=\frac{1}{m\omega}$ remains constant. 

For a Gaussian wave packet $(U=\frac{\hbar^2}{4})$ we find for the uncertainties (\ref{uOszi}) and the chirp (\ref{properties}) using  the abbreviation $k=\frac{K}{m \omega \hbar}$:

\begin{align}
\label{uOsziGauss}
&\Delta_{xp}=\frac{\hbar}{2 }(k^2+1)^{1/2}\,
\mbox{sinh} (2 \omega \tau)\,, \\
&(\Delta x)^2=\frac{\hbar}{2 m \omega}
((k^2+1)^{1/2}\,
\mbox{cosh} (2 \omega \tau)-k)\nonumber\,, \\
&(\Delta p)^2=\frac{m \omega \hbar}{2}
((k^2+1)^{1/2}\,
\mbox{cosh} (2 \omega \tau)+k)\nonumber\,, \\
&a=\frac{m \omega}{2 \hbar} 
\frac{\mbox{sinh}(2 \omega t)}{\mbox{cosh}(2 \omega t)-\frac{k}{\sqrt{k^2+1}}} \quad. \nonumber
\end{align}
\begin{figure}[htp]
\centering
\subfigure[The variable $\tilde a \equiv\frac{a  2 \hbar}{m\omega}$ for $k=-1$ (light-gray),
 $k=0$ (gray) and $k=0.8$. For all three kinds of time evolution $\tilde a$ converges to $\pm 1$  for $k\rightarrow \pm \infty$.]
{\label{uOszia} 
\psfrag {a} {$\tilde a$}
\psfrag{tau}{$\tau$}
\includegraphics[width=0.45\textwidth] {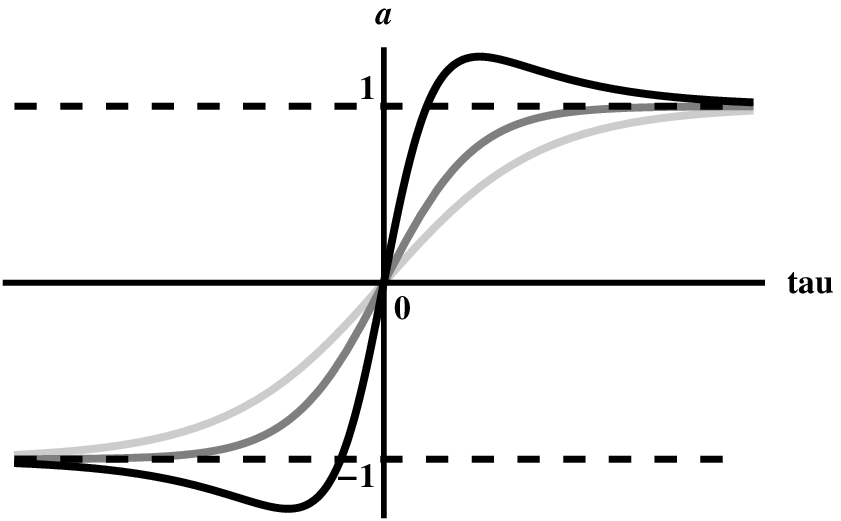}}
\subfigure[The time dependence of $(\Delta X)^2\equiv(\Delta x)^2 \frac{2 m\omega}{\hbar}$),
  $(\Delta P)^2\equiv(\Delta p)^2 \frac{2} {m \omega \hbar}$ (light-gray) and
  $(\Delta_{XP})\equiv(\Delta_{xp}) \frac{2} {\hbar}$ (gray)  for $k=1$.] {\label{uOszib} 
\psfrag {tau} {$ \tau $}
\psfrag {DX2DP2DXP} [cc][cc] {$(\Delta X)^2, (\Delta P)^2, \Delta_{XP}$}
\includegraphics[width=0.45\textwidth]{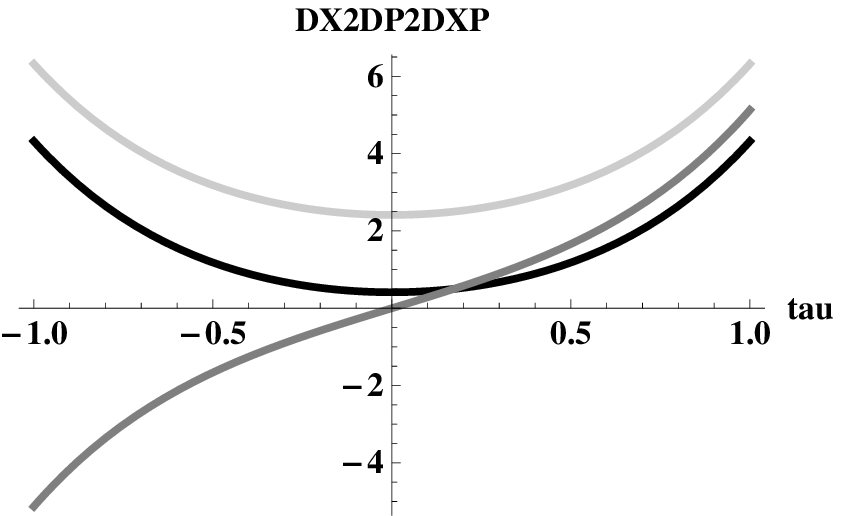}}
\caption{The time evolution of chirp and uncertainties for a Gaussian wave packet in an unstable harmonic oscillator potential.}
\label{uOsziab}
\end{figure}

The parameter $k$ can take values in in the interval $(-\infty,\infty)$. For $k\leq0$
the chirp $a$ is a monotonically increasing function. For $k>0$ the chirp has a maximum and a minimum, respectively, at

\[ 
\tau_{max}= \frac{1}{2 \omega} 
\mbox{arsech} \left( \frac{k}{k^2+1}\right) \quad \mbox{and} \quad
\tau_{min}= -\frac{1}{2 \omega} 
\mbox {arsech} \left( \frac{k}{k^2+1}\right) \qquad,
\]
with the values $a_{max}=\frac{m \omega}{2 \hbar} \sqrt{1+k^2}$
and $a_{min}=-\frac{m \omega}{2 \hbar} \sqrt{1+k^2}$.
If $k$ tends to infinity the interval between $\tau_{max}$ and $\tau_{min}$
 tends to zero. For all possible histories the chirp tends to 
 $\frac {m \omega}{2 \hbar}$ as $\tau$ goes to infinity
 (see also figure \ref{uOszia}).

\section{Conclusions}

All potentials have in common that the product of uncertainties $P=(\Delta x)^2 (\Delta p)^2$ takes the value of its global minimum $U$
 at least once during each possible time evolution (\ref{result1},\ref{result2a},\ref{result2b},\ref{result3}). Furthermore for Gaussian wave packets
 the interval  between the  maximal and minimal  chirp tends  to zero, if the initial value of the constant of motion $K$ tends to infinity.

The harmonic oscillator potential admits constant values for all three uncertainties, if only $(\Delta x_{0})^2\,\geq\, \frac{\hbar}{2 m \omega}$. 
In all other cases the uncertainties oscillate with double frequency compared to the original oscillator frequency.
 This ensures, that the shape of the wave packet does not depend on the direction in which the position expectation value moves.

For all discussed potentials which allow unbounded motion of the position expectation value, we find that wave packets with
 arbitrary initial values are going to spread after a finite time. Furthermore all histories of the position uncertainty consist of a narrowing
 and a spreading part.

Considering these results, it seems worthwhile to pursue the following questions :

Does the fact, that the time-evolved product of uncertainties $P$ has a minimum for all states also hold for arbitrary potentials?

Does any potential that has a stable fixed point admit constant values of uncertainties for certain initial values?

Will all wave packets which evolve in potentials that admit unbounded motion also spread after a certain time?   

\section*{Acknowledgments}

I thank Helmut Rumpf for his careful reading of this article and his valuable advices.


\begin{thebibliography}{99}
              
 \bibitem{He}  E. Heller: `Semiclassical wave packets' in `The Physics and Chemistry of wave packets', Wiley 2000
 \bibitem{Sty} D. Styer: The motion of wave packets through their expectation
                        values and uncertainties, Am. J. Phys. 58(8), 742-744 
(1990)
 \bibitem{An} M. Andrews: Invariant operators for quadratic Hamiltonians, 
                         Am. J. Phys. 67(4), 336-334 (1999) 
 \bibitem{Ali} S. Twareque Ali et al.: Coherent states and their generalizations,
                Reviews in Mathematical Physics, Vol 7, No7 (1995), 1013-1104
   \bibitem {Henry}  R. Henry, S. Glotzer: A squeezed-state primer, Am. J. Phys. 
56(4), 318-328 (1988)  
                                         
 \end{thebibliography}
\end{document}